# Experiences with Permanent Magnets at the Fermilab Recycler Ring

**James T Volk**[a]

[a] *Fermilab National Accelerator Laboratory,
PO Box 500 Batavia Illinois 60510 , USA*

E-mail: volk@fnal.gov

In order to achieve higher luminosities in Run II a separate anti proton storage ring was built in the Main Injector tunnel. To reduce both construction and operations costs permanent magnets were used. This paper discusses the design criterion and specifications, including temperature dependence, longitudinal uniformity, and adjusting of the higher harmonics of the magnets. The design tolerances for a storage ring are more stringent than for a single pass beam line. The difference between the measured and ideal central field for each magnet was held to better than one part in $10^3$. The temperature stability for all magnets was set to better than 1 part in $10^4$ per degree Celsius. Higher order harmonics relative to the central field were set to less than 1 part in $10^4$. This was done for all 484 permanent magnet that were built.

# Contents

## 1. Introduction

The anti proton accumulator at Fermilab was designed to collect and cool anti protons for use in the Fermilab collider. At anti proton stacks on the order of $200 \times 10^{10}$ particles the efficiency fall off and stacks larger than $250 \times 10^{10}$ are not possible. To achieve higher luminosities an alternate method of storing anti protons was designed. A storage ring inside the Main Injector tunnel was proposed electromagnets were considered but permanent magnets low construction costs and minimal operations costs made this the obvious solution. It was originally assumed that anti protons from an existing store would be de-accelerated, re cooled and put into this storage ring hence the name Recycler was chosen. It turns out that anti proton production with this storage ring is sufficient that recycler anti protons is not necessary.

A storage ring has more stringent requirements on field stability, magnetic uniformity and small harmonics than other beam lines or accelerators. It is possible to build permanent magnets that meet the entire criterion necessary to store anti protons for several hundreds of hours and to achieve intensities on the order of $600 \times 10^{10}$ anti protons. Specifications for the magnets and solutions to the construction problems are presented.

## 2. Design criterion

The Recycler was designed as a fixed 8 GeV kinetic energy antiproton storage ring [1]. It is the first ever large scale (3.3 km circumference) accelerator for high energy physics research built with permanent magnets instead of usual electromagnets. It is located in the Main Injector tunnel directly above the Main Injector beamline, near the ceiling, and consists of 362 gradient dipole magnets, and 109 quadrupoles, 8 mirror magnets, and 5 Lambertson magnets. There are four types of gradient magnets: focusing and defocusing, with pole lengths of 3.1 and 4.5 meters. Each has different quadrupole and sextupole fields (see Table 1). The basic design consists of precision shaped pole tips and pole spacers with strontium ferrite bricks on the outside of the poles [2, 3]. The entire package is surrounded by 19 mm (0.75 inch) thick flux return. A typical cross section is shown in Figure 1. A standard SrFe brick size of 101.6 mm by 152.4 mm by 25.4 mm (4 inch by 6 inch by 1 inch) was used. The field axis was orientated along the one inch thickness. Two bricks were used on each side of the pole pieces to generate the required field.

**Table 1: Recycler gradient magnet parameters**

| Magnet | Central Filed | Bdl | Quadrupole | Sextapole |
| --- | --- | --- | --- | --- |
|  | kG | kG-m | kG/m | kG/m$^2$ |
| RGF | 1.3752 | 6.182 | 3.355 | 3.71 |
| RGD | 1.3752 | 6.183 | -3.238 | -6.42 |
| SGF | 1.330 | 4.121 | 6.682 | 0.00 |
| SGD | 1.330 | 4.121 | -6.824 | 0.00 |

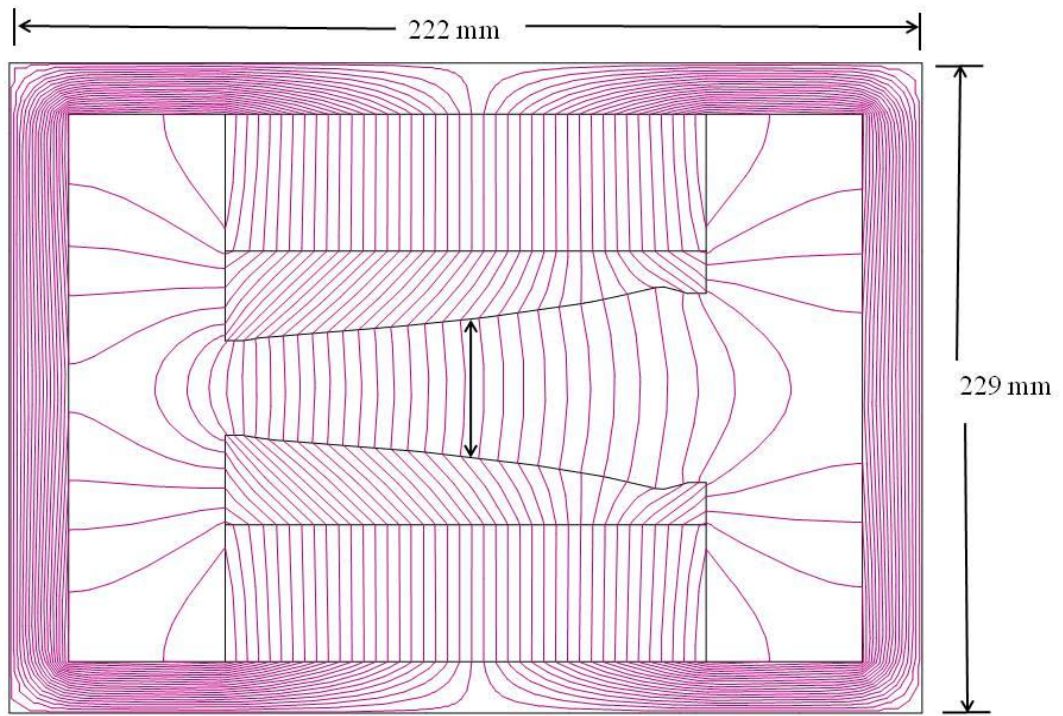

Figure 1: Cross section of the Recycler ring gradient dipole magnet. Magnet gap at center is 50.8 mm high by 152.4 mm wide. Each of the magnet types was modeled using PANDRIA code from LANL.

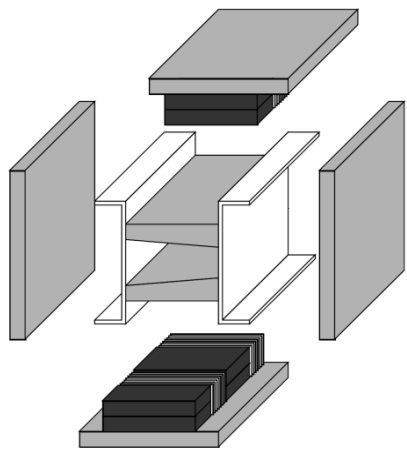

Figure 2: Recycler permanent magnet gradient dipole components shown in an exploded view. For every 4" wide brick there is an 0.5" interval of temperature compensator material composed of 10 strips.

## 3. Magnet Stability

### 3.1 Temperature dependence

The variation of the magnetic field of Strontium ferrite with temperature [4] was carefully measured by freezing the magnet to 0 C then heating the magnet to 40 °C and allowing it to cool. During this process the magnetic field was continually measured with a Morgan (rotating) coil. A least square fit to the data gave the temperature variation *B*-field coefficient of about -0.2% per °C. This intrinsic temperature coefficient of the ferrite material is canceled by interspersing a "compensator alloy" between the ferrite bricks above and below the pole tips – see Fig.2. The compensator is an iron-nickel alloy Ni(30%) Fe (70%) with a low Curie temperature in the range of 40 to 45 °C and therefore its permeability depends strongly on temperature. Thin strips of the alloy shunt away flux in a temperature dependent manner which can be arranged to null out the temperature dependence of the ferrite. The degree of temperature compensation is linearly related to the amount of compensator material in the magnet. Thus the degree of compensation can be "fine tuned" to the required accuracy by adjusting the amount of compensator at the ends of the magnet in a manner similar to the strength trimming with the ferrite. For example, a 20-fold reduction of the temperature coefficient (from 0.2%/°C to 0.01%/°C) requires that the amount of compensator in the magnet be adjusted correctly to 1 part in 20. This poses no difficulties for production. Due to the variable nature of the Curie point from heat to heat, the steel strips from different heats were randomly mixed. All Recycler magnets were frozen to 0 °C during the manufacturing process. The field for each magnet was measured at 0 °C and at 22 °C the amount of compensator steel was adjusted to keep the magnetic field in the 0.01% tolerance. So, as the result the average magnetic field temperature stability of the Recycler magnets was about 0.004%/°C rms over the operating range of 24 to 38 °C – see Fig.3.

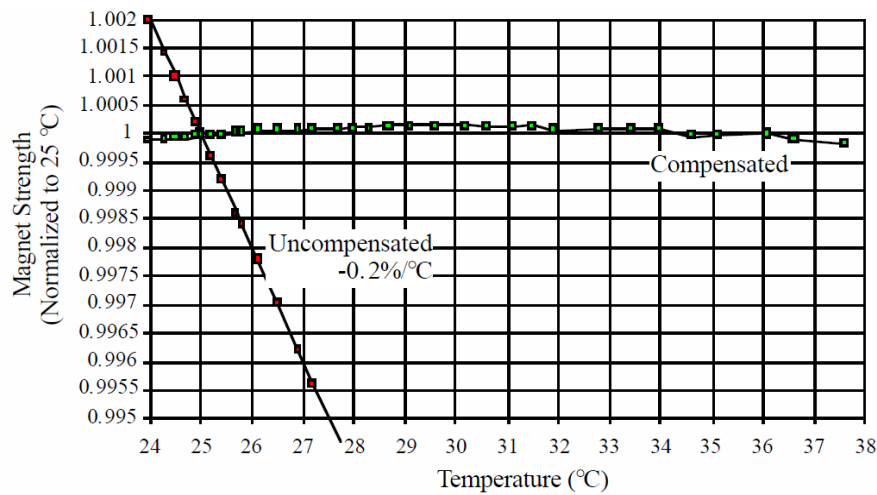

Figure 3 Field strength vs. temperature for a stability test magnet using pre-production samples from the selected vendors for compensator alloy and ferrite.

## 3.2 Time dependence

Long-term field stability was studied using the first production gradient magnet [5]. This magnet was measured with a Morgan (rotating) coil once per week for 12 months during the construction of all the magnets; it was then measured once per month for the next 12 months, then every 6 months for the next ten years. Data were fit to a logarithmic decay – see Figure 4. The magnet lost a total of 0.35% of initial strength in the first 3 years and has remained stable. The fact that over the more than six years of the Recycler operations, its RF frequency of some 52 MHz has changed by less than 200 Hz is further proof that the magnetic field has remained stable within 0.04%.

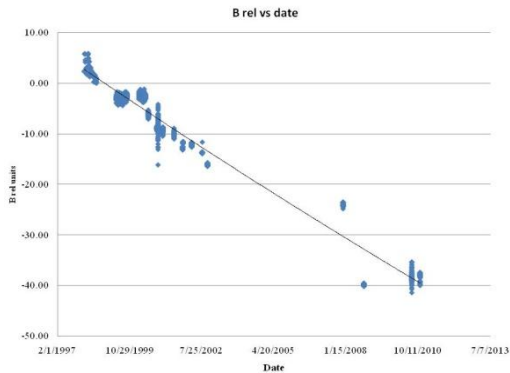

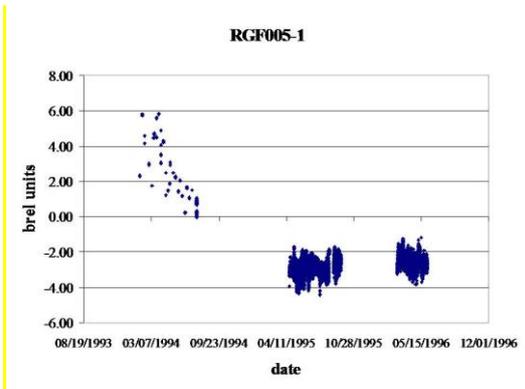

Figure 4: Time evolution of the magnetic field in the Recycler gradient magnet (in the units of $10^{-4}$ of the main dipole field).

## 3.3 Longitudinal uniformity

The longitudinal field for each gradient magnet was measured during production. It was determined that the field could be made more uniform by varying the number of compensator strips between each set of permanent magnets. More compensator strips were added at the quarter points of the magnet and fewer at the ends and middle. This served to keep the field flatter across the length of the magnet and to maintain the temperature compensation within the specified value of 0.01% per °C.

## 3.4 Azimuthal magnetic field harmonics

To deal with the variation of the azimuthal magnetic field harmonics inevitable due to construction imperfections, each gradient magnet was measured with a Morgan (rotating) coil.

Harmonics up to the 12-pole were measured. A FORTRAN program was written to calculate the shape of an end shim for each magnet that would adjust the quadrupole, skew quadrupole, sextupole, and octupole [6]. The program generated machine code for a wire EDM (Electric Discharge Machine) that cut the exact shape required for each magnet. The shims were ready within 24 hours after the first measurement. In general one iteration was required for each magnet. In rare cases two iterations were needed.

The poles of the Recycler quadrupole magnets [7] were made of type 1008 carbon steel. The hyperbolic shape of the poles were formed by cold extruding, no other machining of the poles were required. Strontium ferrite bricks were placed behind the poles. In the corners steel washers were added to adjust the gradient, the skew quad, sextupole and octupole field components. Steel washers were attached to the face of the poles to eliminate the 12-pole component.

Table 2: Quadrupole Magnet Parameters

| Magnet | Gradient kG/m | Focal length 1/m | Gradient * length kG |
|--------|---------------|------------------|----------------------|
| RQMF   | 26.27         | 0.045            | 13.35                |
| RQMD   | -25.32        | -0.043           | -12.86               |
| RQME   | -21.97        | -0.038           | -11.16               |
| RQEB   | 18.46         | 0.063            | 18.76                |
| RQTF   | 16.68         | 0.029            | 8.47                 |
| RQTD   | -16.87        | -0.029           | -8.57                |

## 4. Special magnets

Special magnets needed for the beam injection and extraction - Lambertsons magnets and mirror magnets - were also made with strontium ferrite [8]. The base of the Lambertson magnets was made of solid steel the hole for the field free region was gun bored then through entire 4 meter length. This provided for a field free region of less than 5 Gauss with a field region of 0.16 Tesla. The use of solid base plates allowed for a common design for all 5 Lambertson magnets but different configuration of the bending field.

With experience of operation it became clear that small corrector magnets such as skew quadrupoles and sextapoles were required. Several of these magnets were fabricated to solve specific problems [9]. Both strontium ferrite and samarium cobalt permanent magnets were used.

## Acknowledgments

There were many people involved in the design, construction and installation of the Fermilab Recycler Ring. Foremost among these were the late Klaus Hallbach who provided help and inspiration regarding permanent magnets. G William (Bill) Foster and Gerry P Jackson were the driving force in the conception, design and construction of the Recycler. Other Fermilab personnel involved were K Bertsche, Bruce B Brown, Charles Brown, W B Fowler, David Harding, C Gattuso, Henry Glass, David Johnson, Mike May, C S Mishra, Tom Nicole, J-F Ostiguy, Stan Pruss, Phil Schlabach, M-J Yang.